\documentclass[12pt]{article}
 
\newlength{\dinwidth} \newlength{\dinmargin}
\def\beq{\begin{equation}}
\def\eeq{\end{equation}}
\def\beqa{\begin{eqnarray}}
\def\eeqa{\end{eqnarray}}

\begin{document}
 
\begin{center}
{\Large \bf Collinear and soft resummation in the large-$x$ 
limit}\footnote{Presented at DPF 2004, Riverside, California,
August 26-31, 2004.}
\end{center}
\vspace{2mm}

\begin{center}
{\large Nikolaos Kidonakis}\\
\vspace{2mm}
{\it Kennesaw State University, 1000 Chastain Rd., \#1202\\
Kennesaw, GA 30144-5591, USA\\ and\\
Cavendish Laboratory, University of Cambridge, Madingley Road\\
Cambridge CB3 0HE, England}
\end{center}
  
\vspace{3mm}

\begin{abstract}
I discuss general unified formulas for resumming collinear and soft 
contributions to QCD hard scattering cross sections at large $x$. 
Expansions of the resummed cross sections to next-to-next-to-next-to-leading 
order are also shown along with applications of the formalism. 
\end{abstract}

\thispagestyle{empty} \newpage \setcounter{page}{2}

\section{Introduction}

The calculation of cross sections in perturbative QCD involves the
convolution of hard-scattering factors, which are perturbatively calculable, 
with parton distribution functions.
Near threshold for production of any given system of particles
the restricted phase space for real gluon emission results in large logarithms.
These soft and collinear corrections \cite{SCT}
appear in the form of logarithmic 
plus distributions, 
${\cal D}_l(x)\equiv[\ln^l(1-x)/(1-x)]_+$, where $x \rightarrow 1$ 
at threshold.
There are also purely collinear corrections of the form $\ln^l(1-x)$.

If we define moments of the cross section by
$\hat{\sigma}(N)=\int dx \, x^{N-1} {\hat\sigma}(x)$
then the corrections take the form
$[\ln^{2n-1}(1-x)/(1-x)]_+$ $\rightarrow \ln^{2n}N$
and $\ln^{n}(1-x)$ $\rightarrow \ln^{n}N/N$.
We can formally resum these logarithms in moment space to all orders 
in $\alpha_s$ by
factorizing soft and collinear gluons from the hard 
scattering \cite{KS,NK}.
 
Inverting back to momentum space and expanding the resummed cross section 
at fixed order, we find at NLO soft and collinear corrections 
involving ${\cal D}_1(x)$ and ${\cal D}_0(x)$ terms. 
At NNLO, we have ${\cal D}_3(x)$, ${\cal D}_2(x)$, ${\cal D}_1(x)$,
and ${\cal D}_0(x)$ terms. 

 A unified approach and a master formula for calculating
these distributions at NNLO for any process in hadron-hadron and lepton-hadron 
colliders, for both total and differential cross sections, with simple or
complex color flows, in 1PI or PIM kinematics, and $\overline{\rm MS}$
or DIS factorization schemes, was presented in Ref. \cite{NKuni}.
Results are now available for many processes, including
top pair \cite{NKRV} and FCNC single-top \cite{NKAB} production, 
(charged) Higgs \cite{cHiggs} production,
$W$-boson \cite{NKASV} and direct photon \cite{NKJO,SV} production,
and jet production \cite{KOjets}.

Regarding the purely collinear corrections, at  
NLO we have  $\ln(1-x)$ and constant terms, while at
NNLO we have $\ln^3(1-x)$, $\ln^2(1-x)$, $\ln(1-x)$, and constant terms.
 
\section{Soft and collinear resummation: A unified approach}

The resummed cross section in moment space takes the form \cite{NKuni}
\beqa
{\hat{\sigma}}^{res}(N) &=&   
\exp\left[ \sum_i E_i(N)+E_i^{coll}(N)\right] \; 
\exp\left[ \sum_j E'_j(N)+{E'}_j^{coll}(N)\right] 
\nonumber\\ && \hspace{-20mm} \times \,
\exp \left[\sum_i 2\int_{\mu_F}^{\sqrt{s}} {d\mu' \over \mu'}
\left(\frac{\alpha_s(\mu'^2)}{\pi}\gamma_i^{(1)}
+{\gamma'}_{i/i}(\mu'^2)\right)\right]  
\times
\exp\left[2\, d_{\alpha_s} \int_{\mu_R}^{\sqrt{s}}\frac{d\mu'}{\mu'} 
\beta(\mu'^2)\right] 
\nonumber\\ && \hspace{-20mm} \times 
{\rm Tr} \left \{H(\mu_R^2)
\exp \left[\int_{\sqrt{s}}^{{\sqrt{s}}/{\tilde N}} 
{d\mu' \over \mu'} {\Gamma}_S^\dagger(\mu'^2)\right] 
S (s/{\tilde N}^2) 
\exp \left[\int_{\sqrt{s}}^{{\sqrt{s}}/{\tilde N}} 
{d\mu' \over \mu'} {\Gamma}_S(\mu'^2)\right] \right\}, 
\label{resHS}
\eeqa
where $i$ and $j$ denote incoming and outgoing massless partons, respectively.
In the $\overline{\rm MS}$ scheme, which we will use here,
\beq
E_i(N)=
-\int^1_0 dz \frac{z^{N-1}-1}{1-z}\;
\left \{\int^{\mu_F^2}_{(1-z)^2s} \frac{d\mu'^2}{\mu'^2}
A_i\left(\alpha_s({\mu'}^2)\right)
+{\nu}_i\left[\alpha_s((1-z)^2 s)\right]\right\} 
\nonumber\\
\eeq
with $A_i(\alpha_s) = C_i \, [ {\alpha_s/\pi}+({\alpha_s/\pi})^2 K/2]+A_i^{(3)}
+\cdots$, 
${\nu}_i=(\alpha_s/\pi)C_i+(\alpha_s/\pi)^2 {\nu}_i^{(2)}
+\cdots$
and, for any massless final-state partons at lowest order, 
\beq
E'_j(N)=
\int^1_0 dz \frac{z^{N-1}-1}{1-z}
\left \{\int^{1-z}_{(1-z)^2} \frac{d\lambda}{\lambda}
A_j(\lambda s)-B'_j((1-z)s)
-{\nu}_j((1-z)^2 s)\right\}
\nonumber
\eeq
where $B'_j=(\alpha_s/\pi){B'}_j^{(1)}+(\alpha_s/\pi)^2 {B'}_j^{(2)}+\cdots$
with ${B'}_q^{(1)}=3C_F/4$ and ${B'}_g^{(1)}=\beta_0/4$.

The collinear exponents for incoming and outgoing massless partons are, 
respectively,\\ 
$E_i^{coll}(N)=
\int^1_0 dz \; z^{N-1}\;
\{\int^{\mu_F^2}_{(1-z)^2s} (d\mu'^2/{\mu'^2})
A_i(\alpha_s({\mu'}^2))+\cdots\}$ and\\ 
${E'}_j^{coll}(N)=
-\int^1_0 dz \; z^{N-1}\;
\{\int^{1-z}_{(1-z)^2} (d\lambda/{\lambda})
A_j(\alpha_s(\lambda s))+\cdots\}$.

The $\gamma$'s are parton anomalous dimensions,
$H$ are hard scattering matrices, and 
$S$ are soft matrices that describe noncollinear soft-gluon emission
whose evolution is given by the soft anomalous dimension 
matrices $\Gamma_S$ \cite{KS,NK,NKuni}.

If we expand the resummed cross section, Eq. (\ref{resHS}), to fixed order,
and then invert to momentum space, we get master formulas for the NLO, NNLO, 
and higher-order corrections.
At NLO the master formula for soft and collinear corrections 
is \cite{NKuni}
\beq
{\hat{\sigma}}^{(1)} = \sigma^B \frac{\alpha_s}{\pi}
\left\{c_3 {\cal D}_1(x) + c_2  {\cal D}_0(x) \right\}
+\frac{\alpha_s^{d_{\alpha_s}+1}}{\pi} A^c  {\cal D}_0(x)\, 
\eeq
where $\sigma^B$ is the Born term, $\alpha_s$ is at scale $\mu_R$, 
$c_3=\sum_i 2 \, C_i -\sum_j C_j$,
with $C_F=(N_c^2-1)/(2N_c)$ for quarks  
and $C_A=N_c$ for gluons,
\beq
c_2=\! - \!\sum_i \! \left[C_i
+C_i \ln\left(\frac{\mu_F^2}{s}\right)\right]
-\sum_j \! \left[{B'}_j^{(1)}+C_j
+C_j \ln\left(\frac{M^2}{s}\right)\right]\, , 
\eeq
with $M$ a hard scale, and
$A^c={\rm tr} (H^{(0)} {\Gamma}_S^{(1)\,\dagger} S^{(0)}
+H^{(0)} S^{(0)} {\Gamma}_S^{(1)})$.
We can also calculate the purely collinear terms from the expansion. 
As an example, for the Drell-Yan process, $q {\bar q} \rightarrow V$, 
whose cross section is known to NNLO \cite{HvNM}, we find
\beq
{\hat{\sigma}}^{(1)}=\sigma^B \frac{\alpha_s}{\pi}
\left\{4 C_F \left[\frac{\ln(1-x)}{1-x}\right]_+ \! \!
-2C_F \ln\left(\frac{\mu_F^2}{Q^2}\right) \left[\frac{1}{1-x}\right]_+ 
\! \! -4C_F \ln(1-x)\right\}.
\eeq

At NNLO the master formula for the soft and collinear corrections
is \cite{NKuni}
\beqa
{\hat{\sigma}}^{(2)}&=& 
\sigma^B \frac{\alpha_s^2(\mu_R^2)}{\pi^2} \,
\frac{1}{2} c_3^2 \, {\cal D}_3(x)
+\sigma^B \frac{\alpha_s^2(\mu_R^2)}{\pi^2} 
\left\{\frac{3}{2} c_3  c_2 - \frac{\beta_0}{4}  c_3
+\sum_j C_j \, \frac{\beta_0}{8}\right\} \, {\cal D}_2(x)
\nonumber \\ && 
{}+\frac{\alpha_s^{d_{\alpha_s}+2}(\mu_R^2)}{\pi^2} \,
\frac{3}{2} \, c_3 \, A^c\, {\cal D}_2(x) +\cdots
\eeqa
where here we show only the leading and next-to-leading logarithms.
We note that two-loop calculations \cite{NKtwoloop} are needed to get 
all soft logarithms at NNLO.

We can continue this procedure to higher orders.
At next-to-next-to-next-to-leading order (NNNLO) the master formula
is
\beqa
{\hat{\sigma}}^{(3)}&=& 
\sigma^B \frac{\alpha_s^3(\mu_R^2)}{\pi^3} \;  
\frac{1}{8} \, c_3^3 \; {\cal D}_5(x)
\nonumber \\ && \hspace{-20mm}
{}+\sigma^B \frac{\alpha_s^3(\mu_R^2)}{\pi^3} \; 
\left\{\frac{5}{8} \, c_3^2 \, c_2 -\frac{5}{2} \, c_3 \, X_3\right\} \;  
{\cal D}_4(x)
+\frac{\alpha_s^{d_{\alpha_s}+3}(\mu_R^2)}{\pi^3} \;
\frac{5}{8} \, c_3^2 \, A^c \; {\cal D}_4(x) +\cdots
\eeqa
where 
$X_3=(\beta_0/12) c_3-\sum_j C_j \beta_0/24$,
and again we only show explicitly the leading and next-to-leading logarithms.


\begin{thebibliography}{0}
 
\bibitem{SCT}
G. Sterman, {\it Nucl. Phys.} {\bf B281}, 310 (1987).

\bibitem{KS}
N. Kidonakis and G. Sterman, {\it Phys. Lett.} {\bf B387}, 867 (1996);
{\it Nucl. Phys.} {\bf B505}, 321 (1997);
N. Kidonakis, G. Oderda, and G. Sterman,
{\it Nucl. Phys.} {\bf B531}, 365 (1998).
 
\bibitem{NK}
N. Kidonakis, {\it Int. J. Mod. Phys.} {\bf A15}, 1245 (2000);
{\it Mod. Phys. Lett.} {\bf A19}, 405 (2004).
 
\bibitem{NKuni}
N. Kidonakis, {\it Int. J. Mod. Phys.} {\bf A19}, 1793 (2004);
in DIS03, hep-ph/0306125, hep-ph/0307207.
 
\bibitem{NKRV}
N. Kidonakis, {\it Phys. Rev.} {\bf D64}, 014009 (2001);
N. Kidonakis and R. Vogt, {\it Phys. Rev.} {\bf D68}, 114014 (2003);
{\it Eur. Phys. J.} {\bf C33}, s466 (2004).

\bibitem{NKAB}
A. Belyaev and N. Kidonakis, {\it Phys. Rev.} {\bf D65}, 037501 (2002);
N. Kidonakis and A. Belyaev, {\it JHEP} {\bf 12}, 004 (2003);
in DIS04, hep-ph/0407032.

\bibitem{cHiggs}
The Higgs Working Group, hep-ph/0406152;
N. Kidonakis, in DIS04, hep-ph/0406179.

\bibitem{NKASV}
N. Kidonakis and A. Sabio Vera, {\it JHEP} {\bf 02}, 027 (2004);
in DIS04, hep-ph/0405013; in DPF04, hep-ph/0409206;
hep-ph/0409337.
 
\bibitem{NKJO}
N. Kidonakis and J.F. Owens, {\it Phys. Rev. D} {\bf 61}, 094004 (2000);
in EPS-HEP99, hep-ph/9910240;
{\it Int. J. Mod. Phys.} {\bf A19}, 149 (2004).

\bibitem{SV}
G. Sterman and W. Vogelsang, {\it JHEP} {\bf 02}, 016 (2001).

\bibitem{KOjets}
N. Kidonakis and J.F. Owens, {\it Phys. Rev. D} {\bf 63}, 054019 (2001).

\bibitem{HvNM}
R. Hamberg, W.L. van Neerven, and T. Matsuura,
{\it Nucl. Phys.} {\bf B359}, 343 (1991).

\bibitem{NKtwoloop}
N. Kidonakis, hep-ph/0208056; in DIS2003, hep-ph/0307145.

\end{thebibliography}
\end{document}